\documentclass[aps,prl,reprint,superscriptaddress]{revtex4-2}
\usepackage{amsmath}
\usepackage{amssymb}
\usepackage{graphicx}
\usepackage{xcolor}
\usepackage{CJKutf8}
\usepackage[colorlinks=true, linkcolor=blue, citecolor=blue, urlcolor=blue]{hyperref}
\usepackage{orcidlink}

\newcommand{\COMMENTED}[1]{}

\definecolor{pyang}{rgb}{0.83, 0.69, 0.22}

\newcommand{\bs}{\boldsymbol}
\graphicspath{{./figures/main/}}

\begin{document}

\title{Singlet--Triplet Transition of Rotating Moir\'e Atoms in a Magnetic Field}

\author{Conor Smith \orcidlink{0009-0005-4580-762X}}
\affiliation{Center for Computational Quantum Physics, Flatiron Institute, New York, NY, 10010, USA}
\author{Yubo Yang (\texorpdfstring{\begin{CJK*}{UTF8}{gbsn}杨煜波\end{CJK*}}{Yubo Yang}) \orcidlink{0000-0002-8800-9426}}
\affiliation{Center for Computational Quantum Physics, Flatiron Institute, New York, NY, 10010, USA}
\affiliation{Department of Physics and Astronomy, Hofstra University, Hempstead, NY, 11549, USA}
\author{Shiwei Zhang \orcidlink{0000-0001-9635-170X}}
\affiliation{Center for Computational Quantum Physics, Flatiron Institute, New York, NY, 10010, USA}
\author{Miguel A. Morales \orcidlink{0000-0002-6389-3067}}
\affiliation{Center for Computational Quantum Physics, Flatiron Institute, New York, NY, 10010, USA}

\date{\today}

\begin{abstract}
Moir\'e superlattices can bind several electrons to each potential minimum, forming few-electron ``moir\'e atoms.'' Using neural quantum states that obey magnetic periodic boundary conditions, include full Landau-level mixing, and treat both spin sectors, we map the ground state of two-electron moir\'e atoms in a superlattice as a function of moir\'e depth and magnetic field strength. At strong field each atom is a spin-triplet maximum-density droplet with canonical angular momentum $\langle L_z\rangle=-1$ and a persistent circulating current: the rotating counterpart of the ``Wigner molecular crystal'' recently imaged in moir\'e superlattices. Deepening the wells drives a singlet--triplet transition to non-rotating spin-singlet atoms whose only current is a weak diamagnetic counter-circulation. The transition moves to deeper wells as the field grows, so depth and field together open a window for the formation of rotors. This is the collective analog of the singlet--triplet transition of two-electron quantum dots. At the transition, the orbital magnetic moments of the atoms reverse sign as their spin polarization switches, a joint signature accessible to magnetometry. If the atoms are instead polarized by other means, such as magnetic proximity coupling, the rotors lock at weak field into a spontaneously ordered crystal of dimers in a pinwheel pattern.
\end{abstract}

\maketitle

\emph{Introduction.}---Moir\'e superlattices of transition-metal dichalcogenides host generalized Wigner crystals~\cite{Regan2020,Xu2020,Li2021,Huang2021,Jin2021}, fractional Chern insulators~\cite{Cai2023,Zeng2023,Park2023,XuF2023}, and fractional quantum spin Hall states~\cite{Kang2024}. When there are more electrons than potential minima, several electrons bind to each minimum and the resulting ``moir\'e atoms'' form Wigner molecules~\cite{ReddyDevakulFu2023,YannouleasLandman2023}, whose ordered arrays, termed Wigner molecular crystals, have been imaged directly~\cite{li2024wigner} and studied with neural-network wavefunctions~\cite{Li2025,PWC}. A perpendicular magnetic field adds a new energy scale. In an isolated two-electron quantum dot it drives a transition from a spin singlet, with both electrons in the lowest orbital, to a spin triplet that carries one quantum of angular momentum~\cite{Wagner1992,MaksymChakraborty1990,ReimannManninen2002}, seen directly in capacitance spectroscopy~\cite{Ashoori1993} and in vertical dots~\cite{Tarucha1996,Kouwenhoven2001}. A transition of this kind has also been seen in a moir\'e lattice where local compressibility in twisted WSe$_2$/MoSe$_2$ finds singlets at doubly occupied sites giving way to triplets as the field is raised~\cite{Kometter2023}. What such a measurement cannot resolve is the orbital state. The triplet differs from the singlet not only in spin but also in the angular momentum, which Pauli exclusion forces it to carry. Whether that circulation survives the Coulomb coupling between atoms and what it does to the order between them is an open question. The moir\'e lattice hosts strong interactions between neighboring atoms, allowing the possibility of a collective orientation and crystallization.

The confinement set by the twist angle and the magnetic field strength are independent and experimentally tunable so the competition between the two can be studied in a single device. 
This provides an opportunity for examining and understanding emergent quantum phenomena as a result of competition between magnetic field, band structure, and strong electron interaction.
However, it has been highly challenging to study this competition accurately with many body methods. The moir\'e depth, the cyclotron energy, and the Coulomb repulsion are all comparable. Neither projection to a single Landau level nor a mean-field treatment of the spin sectors is adequate to accurately model this competition. Here we address this challenge with a neural quantum state (NQS) approach, which
has been applied successfully to the zero-field Wigner phases of moir\'e materials~\cite{Li2025,PWC} but so far applications to magnetic fields on the torus have been limited to small system sizes and without moir\'e potentials~\cite{MagNet,PerezFadon2025}. We address this gap with an NQS that satisfies magnetic boundary conditions while retaining full Landau-level mixing.

\begin{figure}[t]
\centering
\includegraphics[width=\linewidth]{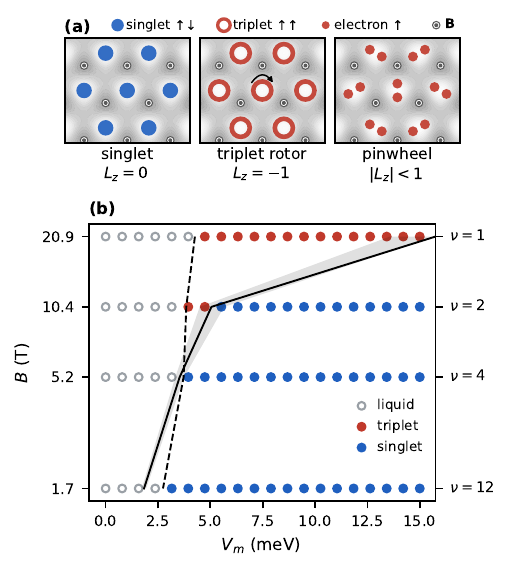}
\caption{(a) Schematic of the three kinds of moir\'e atoms on the $\phi=40^\circ$ moir\'e potential (gray): a spin singlet with both electrons in the lowest orbital ($L_z=0$), a spin-triplet rotor (MDD) with $L_z=-1$ and, at weak field in the polarized sector, a two-lobed dimer whose axis orders into a pinwheel crystal. (b) Ground state versus moir\'e depth $V_M$ and field $B$ (right axis: filling $\nu$), at $r_s=15$ with two electrons per cell. Open circles: liquid (atoms not yet formed; fewer than $1.95$ electrons per deep minimum). Filled symbols: single electron (red), triplet rotor phase (red with white center) and singlet phase (blue), assigned from $E_S-E_T$ including the Zeeman energy (Fig.~\ref{fig:st}). Solid line: singlet--triplet transition $V_M^*(B)$ for $g_0=2.5$, with a shaded band for $g_0=2.1$--$2.9$ (at $\nu=1$ it lies beyond $15$~meV); dashed line: onset of atom formation.}
\label{fig:phase}
\end{figure}

\begin{figure*}[t]
\centering
\includegraphics[width=\linewidth]{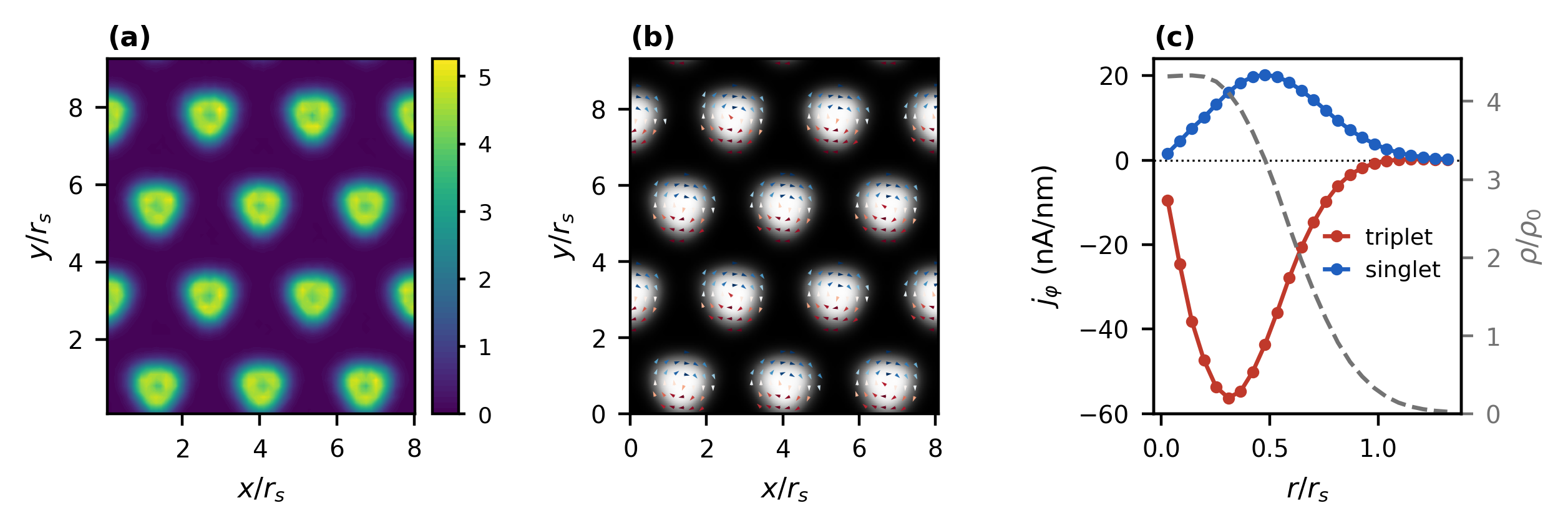}
\caption{Rotating moir\'e atoms at $\nu=1$ ($B=20.9$~T), $V_M=10.3$~meV, polarized sector. (a) Density $\rho/\rho_0$. (b) Gauge-invariant current density $\bs{j}$: arrows are unit vectors along $\bs{j}$ colored by $\hat{j}_x$, drawn where $\rho>12\%$ of its maximum, over $\rho/\rho_0$ in grayscale. (c) Azimuthal component of $\bs{j}$ about the moir\'e minimum, averaged over all atoms [Eq.~(\ref{eq:jphi}); red, left axis], and radial density $\rho(r)/\rho_0$ (gray dashed, right axis). Blue: the same for the unpolarized (singlet) state at this depth, where it is metastable, showing a weaker counter-circulation. The current is plotted as the electric current density $e\,j_\varphi$ in nA/nm, i.e.\ the current crossing each nanometer of a radial cut, with the sign following the electron circulation (negative = clockwise); Eq.~(\ref{eq:Lj}) uses the particle current $j_\varphi$ itself and returns $\langle L_z^{\rm mech}\rangle=-0.41$ for the triplet and $+0.36$ for the singlet. Lengths are in units of $r_s a_B^*=7.9$~nm.}
\label{fig:rotor}
\end{figure*}

\begin{figure}[t]
\centering
\includegraphics[width=\linewidth]{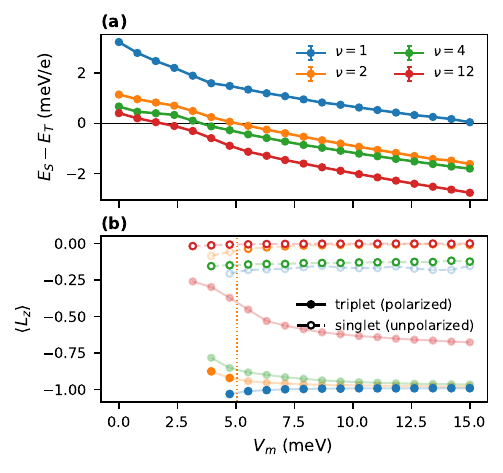}
\caption{Field-driven singlet--triplet transition of the moir\'e atoms. (a) $E_S-E_T$ per electron, the unpolarized minus the polarized energy plus the Zeeman energy ($g_0=2.5$), versus $V_M$ at $\nu=1,2,4,12$. Positive values indicate that the triplet rotor is the ground state. (b) Canonical $\langle L_z\rangle$ per atom, Eq.~(\ref{eq:Lmech}), in the polarized (triplet; filled symbols, solid lines) and unpolarized (singlet; open symbols, dashed lines) sectors. Points are drawn in full where that sector is the ground state and faded where it is metastable; the dotted line marks $V_M^*$ at $\nu=2$. Liquid points are omitted.}
\label{fig:st}
\end{figure}

Here we show that the singlet--triplet transition survives in the presence of Coulomb coupling and that indeed it can serve as an underlying principle that organizes the phase diagram. At strong field, each two-electron atom is a spin-triplet maximum-density droplet (MDD)~\cite{MacDonaldYangJohnson1993,YannouleasLandman2007,Oosterkamp1999} with canonical angular momentum $\langle L_z\rangle=-1$ and a circulating persistent current [Figs.~\ref{fig:phase}(a) and \ref{fig:rotor}]. Deepening the moir\'e wells raises the cost of the orbital excitation the triplet requires, and every atom turns into a non-rotating singlet at a critical depth $V_M^*$ (Fig.~\ref{fig:st}). Because the Zeeman energy grows with the field while the cost of that excitation does not~\cite{supp}, $V_M^*$ increases with field strength and the rotor phase occupies a window bounded below by atom formation and above by $V_M^*$ [Fig.~\ref{fig:phase}(b)]. This is the singlet--triplet transition of a two-electron quantum dot, tuned by confinement rather than by field~\cite{Kyriakidis2002}. In the presence of a weak field, the atoms orient
and spontaneously break the moir\'e symmetry by forming a $\sqrt3\times\sqrt3$ pinwheel crystal shown in the right panel of Fig.~\ref{fig:phase}(a). 
The field therefore sets not only whether each atom is polarized but whether it rotates, and, through that rotation, how the atoms orient with respect to one another. These high accuracy results not only shed light on existing experiment but will help guide a variety of future ones.

\emph{Model.}---We consider electrons doped into the MoSe$_2$ layer of a MoSe$_2$/WSe$_2$ heterobilayer, with effective mass $m^*=0.5\,m_e$~\cite{Kormanyos2015} and permittivity $\epsilon_r=5$~\cite{Laturia2018}, so that $a_B^*=0.53$~nm and $\mathrm{Ha}^*=544$~meV. In a perpendicular field, the Hamiltonian is
\begin{equation}\label{eq:H}
\mathcal{H} = \frac{1}{2}\sum_i \bs{\Pi}_i^2 - \sum_i V(\bs{r}_i) + \sum_{i<j}\frac{1}{|\bs{r}_i-\bs{r}_j|} + \mathcal{H}_Z,
\end{equation}
in effective atomic units, with $\bs{\Pi}_i=-i\bs{\nabla}_i+\bs{A}(\bs{r}_i)$ and, in the Landau gauge, $\bs{A}=B(-y,0)$. The moir\'e potential takes the leading-harmonic form~\cite{WuMacDonald2018} $V(\bs{r})=2V_M\sum_{j=1}^3\cos(\bs{g}_j\cdot\bs{r}+\phi)$ with $\phi=40^\circ$, for which each moir\'e cell has a single deep minimum. We fix two electrons per moir\'e cell at $r_s=15$, which corresponds to a moir\'e period $a_M=21.4$~nm, and vary the depth over $0\le V_M\le15$~meV. The Landau-level filling $\nu=N_e/N_\phi$ sets the field, $B=20.9~\mathrm{T}/\nu$ (Table~\ref{tab:params}). The Zeeman term $\mathcal{H}_Z=\tfrac12 g_0\mu_B B\sum_i\sigma^z_i$ enters only in the comparison between spin sectors. We use band values of $m^*$ and $g$, since their interaction renormalization is contained in the many-body energies: the bare conduction-band $g$-factor $g_0=2.5\pm0.4$ of MoSe$_2$~\cite{MoSe2chi2023}, rather than the enhanced $g^*\sim 7$ to $13$ seen in transport, and the band mass rather than the cyclotron mass $0.8\,m_e$~\cite{Larentis2018}.

\begin{table}[b]
\caption{Parameters at $r_s=15$ with two electrons per moir\'e cell. $\hbar\omega_c$ and $E_Z=\tfrac12 g_0\mu_B B$ (per electron, $g_0=2.5$) are listed at each filling studied.}
\label{tab:params}
\begin{ruledtabular}
\begin{tabular}{lccc}
$\nu$ & $B$ (T) & $\hbar\omega_c$ (meV) & $E_Z$ (meV) \\
\hline
1  & 20.9 & 4.84 & 1.51 \\
2  & 10.4 & 2.42 & 0.76 \\
4  & 5.2  & 1.21 & 0.38 \\
12 & 1.7  & 0.40 & 0.13 \\
\end{tabular}
\end{ruledtabular}
\end{table}

\emph{Method.}---
We study this model by variational Monte Carlo with a neural network based wavefunction. The approach is powerful in this context because it does not require a priori assumptions about 
which Landau levels are occupied or how electrons correlate. The same ansatz describes weakly correlated liquids, a lattice of localized circularly symmetric pairs, and a pinwheel crystal while treating spin sectors on equal footing. All states are obtained with the same optimization machinery and ground states are determined solely by energy minimization.

On a torus threaded by $N_\phi$ flux quanta, the wavefunction satisfies magnetic periodic boundary conditions. Our ansatz is a Slater determinant of Landau-level orbitals mixed by learnable, $V_M$-conditioned coefficients, multiplied by an attention-based neural multiplicative orbital backflow and a Jastrow factor~\cite{Rende_2025,supp}.
Similar to other real-space neural ans\"atze, we treat the quantum Hall problem without projecting to a single Landau level~\cite{MagNet,PerezFadon2025,Teng2025,Qian2025}. This construction obeys the magnetic boundary conditions and is not restricted to the lowest Landau level. A single optimization conditioned on $V_M$ yields the ground state at all depths for a given field and spin sector~\cite{supp}. We optimize the fully polarized sector ($S_z=N_e/2$) and the unpolarized sector ($S_z=0$) separately, for a $3\times4$ cell of $12$ moir\'e atoms ($N_e=24$). The unpolarized state places one electron of each spin in the lowest orbital of every atom; we therefore refer to it as the singlet.

\emph{Observables.}---We sample configurations $R=(\bs{r}_1,\dots,\bs{r}_{N_e})$ from $|\Psi|^2$ and smooth the electron positions with a periodic Gaussian $\delta_\sigma$ of width $\sigma=0.06\,a_M$. The density and the particle current density are
\begin{align}
\rho(\bs{r})&=\Big\langle\sum_i\delta_\sigma(\bs{r}-\bs{r}_i)\Big\rangle,\label{eq:rho}\\
\bs{j}(\bs{r})&=\Big\langle\sum_i\delta_\sigma(\bs{r}-\bs{r}_i)\,\bs{v}_i\Big\rangle,\label{eq:rhoj}
\end{align}
where
\begin{equation}\label{eq:v}
\bs{v}_i=\mathrm{Re}\,\frac{\bs{\Pi}_i\Psi}{\Psi}=\mathrm{Im}\,\bs{\nabla}_i\ln\Psi+\bs{A}(\bs{r}_i)
\end{equation}
is the gauge-invariant velocity of electron $i$; the electric current is $-e\bs{j}$. The canonical and mechanical angular momenta of atom $s$ are
\begin{equation}\label{eq:Lz}
L_z^{\rm can}=\sum_{i\in s}\big(\bs{d}_i\times\bs{p}_i\big)_z,\qquad
L_z^{\rm mech}=\sum_{i\in s}\big(\bs{d}_i\times\bs{\Pi}_i\big)_z,
\end{equation}
with $\bs{\Pi}_i=\bs{p}_i+\bs{A}(\bs{r}_i)$.
The distance between each electron and its atomic center, $d_i$, is computed as follows: each electron is assigned to its nearest deep minimum $\bs{R}_s$ (within $a_M/2$), and we define $\bs{d}_i=\bs{r}_i-\bs{R}_s$ using the minimum-image convention.
We report expectation values averaged over the $N_s=12$ atoms. $L_z^{\rm mech}$ is gauge invariant and can be directly computed. For $L_z^{\rm can}$, $\bs{p}_i=-i\bs{\nabla}_i$ is taken in the symmetric gauge centered on the atom, $\bs{A}=\tfrac{B}{2}\hat{\bs{z}}\times\bs{d}$, in which $L_z^{\rm can}$ coincides with the angular-momentum quantum number of a rotationally symmetric atom. In this gauge $(\bs{d}\times\bs{A})_z=\tfrac{B}{2}|\bs{d}|^2$, so the two differ by the diamagnetic response of the droplet,
\begin{equation}\label{eq:Lmech}
\langle L_z^{\rm mech}\rangle=\langle L_z^{\rm can}\rangle+\frac{B}{2}\sum_{i\in s}\langle|\bs{d}_i|^2\rangle ,
\end{equation}
and we obtain $\langle L_z^{\rm can}\rangle$ from this relation, using the gauge-invariant $L_z^{\rm mech}$.
Finally, we resolve the mechanical angular momentum radially using the component of $\bs{j}$ that circulates about a minimum,
\begin{equation}\label{eq:jphi}
j_\varphi(r)=\frac{1}{N_s}\sum_s\big\langle\bs{j}(\bs{R}_s+\bs{d})\cdot(\hat{\bs{z}}\times\hat{\bs{d}})\big\rangle_{|\bs{d}|=r},
\end{equation}
where $\hat{\bs{z}}\times\hat{\bs{d}}$ is the azimuthal unit vector and the average runs over directions at fixed distance $r$, so that
\begin{equation}\label{eq:Lj}
\langle L_z^{\rm mech}\rangle=2\pi\int_0^{a_M/2}r^2\,j_\varphi(r)\,dr .
\end{equation}

\emph{Rotating atoms.}---Figure~\ref{fig:rotor} shows the ground state at $\nu=1$ and $V_M=10.3$~meV, which is in the triplet phase. Each minimum binds two electrons into a compact, $C_3$-symmetric droplet [Fig.~\ref{fig:rotor}(a)]. The current density circulates clockwise around every atom [Fig.~\ref{fig:rotor}(b)], peaking on the inner flank of the droplet at $j_\varphi\approx-56 \,\text{nA}/\text{nm}$ [Fig.~\ref{fig:rotor}(c)]. The canonical angular momentum locks onto the MDD value of $-1$ as soon as the atoms form, at $V_M\approx4$~meV, and remains there up to the deepest wells [Fig.~\ref{fig:st}(b)]. Pauli exclusion requires non-zero angular momentum since two parallel spins cannot share the $m=0$ orbital, so the pair must carry relative angular momentum, and the field determines its sign. The mechanical angular momentum is smaller because the diamagnetic term in Eq.~(\ref{eq:Lmech}) opposes the rotation and the radius of the droplet is constant as the field is lowered.

\emph{Singlet--triplet transition.}---While a strong magnetic field favors the spin-aligned triplet, a deep moir\'e potential stabilizes the more spatially compact singlet. Figure~\ref{fig:st}(a) compares the two spin sectors as a function of depth at four field strengths, using the energy per electron of the unpolarized state relative to the polarized one, including the Zeeman energy gained by the polarized state,
\begin{equation}\label{eq:EST}
E_S-E_T=\frac{E_{S_z=0}-E_{S_z=N_e/2}}{N_e}+\frac{1}{2}g_0\mu_B B .
\end{equation}
 At $V_M=0$ and $\nu=1$ the system is a quantum Hall liquid and the polarized state is lower in energy, as expected for a quantum Hall ferromagnet. Once atoms form, the competition becomes local and is easiest to follow one atom at a time~\cite{Wagner1992,ReimannManninen2002}. The singlet puts both electrons in the lowest orbital, which opposite spins are allowed to share. Parallel spins, however, cannot occupy the same orbital, so, in the triplet, the second electron must occupy a different orbital; in a magnetic field, the lowest available one is the $m=-1$ angular-momentum state.

Deepening the well raises the cost of the $m=-1$ state
but does not change the Zeeman energy, which only grows with the field. Every atom therefore reverts to a singlet beyond a critical depth $V_M^*$, and a larger field pushes $V_M^*$ deeper~\cite{supp}. At $\nu=1$, the triplet is the ground state from atom formation ($4.3$~meV) out to the deepest wells computed ($15$~meV). Halving the field nearly closes the window and by $\nu=4$ it has closed altogether [Fig.~\ref{fig:phase}(b)]. The Zeeman energy is what holds the window open at $\nu=1$ but only a modest $g$-factor of $g_c=1.45$ is needed.

The window at $\nu=1$ is much wider than at weaker magnetic fields because each Landau level holds one state per flux quantum, so the field determines how many of them each atom is allotted. At $\nu=1$, two flux quanta thread every moir\'e cell, so each atom receives two lowest-Landau-level orbitals and the triplet fits entirely within the lowest Landau level. At $\nu=2$, each atom has only one orbital per Landau level, and the triplet's second electron must cross the Landau gap at a cost of $\hbar\omega_c$ per atom. This, along with the halving of the Zeeman energy, accounts for the step between the curves. The wide rotor window is therefore tied to $\nu\le1$, where the lowest Landau level has room for both orbitals of every atom.

The transition is visible in the angular momentum [Fig.~\ref{fig:st}(b)]. The triplet carries nearly the full quantum of circulation and the singlet very little. For $\nu=2$ the ground state passes from the triplet to the singlet at $V_M^*$. Nothing forces the singlet's electrons out of the $m=0$ orbital, so its canonical part is small and the diamagnetic term of Eq.~(\ref{eq:Lmech}) outweighs it. The pair therefore counter-circulates and the current reverses at the transition [Fig.~\ref{fig:rotor}(c)]. Since the magnetization follows the mechanical angular momentum, $\mu_z=-\tfrac12\langle L_z^{\rm mech}\rangle$~\cite{supp}, the orbital moment of every atom changes sign at $V_M^*$, together with the spin polarization.

\begin{figure}[t]
\centering
\includegraphics[width=\linewidth]{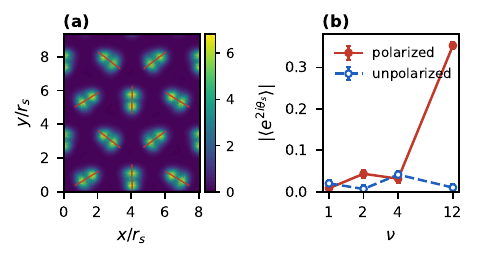}
\caption{Orientational crystal of the polarized sector. (a) Density $\rho/\rho_0$ at $\nu=12$ ($B=1.7$~T), $V_M=10.3$~meV, polarized sector: each atom is a two-lobed dimer. Red bars: mean dimer axis of each atom, taking three orientations $\approx60^\circ$ apart on the sublattices of a $\sqrt3\times\sqrt3$ pinwheel. Lengths in units of $r_s a_B^*$. (b) Orientational order parameter $|\langle e^{2i\theta_s}\rangle|$, where $\theta_s$ is the direction of the relative coordinate $\bs{r}_1-\bs{r}_2$ of the pair on atom $s$, measured from a fixed lattice axis; the angle is doubled because the two ends of a dimer are equivalent. Averaged over atoms and shown versus filling at the same depth, in the polarized (filled) and unpolarized (open) sectors; error bars are block standard errors.}
\label{fig:pinwheel}
\end{figure}

\emph{Polarized sector.}---If the atoms were polarized by other means, such as a larger $g$-factor, there would be no singlet phase to cross into, and the ground state would follow the polarized curve of Fig.~\ref{fig:st}(b) down to arbitrarily weak fields. Along it, the atoms remain circular up to $\nu=4$, but by $\nu=12$ each is a two-lobed dimer [Fig.~\ref{fig:pinwheel}(b)], and the dimer axes order across the lattice into a pinwheel pattern [Fig.~\ref{fig:pinwheel}(a)]. This order is spontaneous, breaks the $C_3$ symmetry of each moir\'e cell, and triples the unit cell. The axes on the three sublattices of a $\sqrt3\times\sqrt3$ superlattice point about $60^\circ$ apart.

A rotating atom and a stationary dimer are built from the same two states. With both spins parallel, Pauli exclusion makes the pair's relative wavefunction odd under exchange, so its angular momentum must be odd and the two lowest such states are the opposite senses of circulation, $L_z=\pm1$. Occupying one of them gives a rotor with a ring-shaped density, whereas an equal superposition gives a standing wave: a dimer pointing along a definite axis. The field separates the two states by $\hbar\omega_c$, so anything that mixes them competes directly with the field. At $\nu=1$, the field wins and the atoms rotate; as $\hbar\omega_c$ decreases, the mixing wins and they lock into the pinwheel orientation.

The mixing is not supplied by the moir\'e potential: its anisotropy about each minimum is three-fold, so for a two-lobed atom one lobe would sit on a minimum of that anisotropy while the other would sit on a maximum~\cite{YannouleasLandman2024PRL}. The mixing arises instead from the interactions between atoms~\cite{YannouleasLandman2024PRB}: minimizing the classical electrostatic energy of two point charges per minimum in the same moir\'e potential yields the $\sqrt3\times\sqrt3$ pinwheel. The axis order parameter is the nematic order parameter of a triangular lattice, studied for moir\'e systems at other fillings~\cite{Matty2022,Jin2021}. What remains is a competition between the field, which drives each atom to circulate, and the Coulomb interaction with neighboring dimers, which drives the atoms to orient collectively.

What the pinwheel requires is that the atoms be polarized at this weak field. The critical $g$-factor rises with depth, and at $\nu=12$ the crossing falls just below the depth at which atoms form~\cite{supp}. A larger valley $g$-factor or proximity to a ferromagnetic insulator~\cite{Zhong2017,Zhao2017} would bring the polarized sector into the range of depths where the dimers exist. We therefore expect the pinwheel to be reachable in realistic devices.

\emph{Discussion.}--- At strong magnetic field, the effect is more significant than polarizing spins. The strength of the field determines whether each atom rotates and, through that, whether the atoms collectively orient into a crystal.  When the lowest Landau level supplies exactly the two orbitals a spin-polarized pair requires, every atom in the lattice becomes a quantum rotor, carrying one quantum of angular momentum and a persistent current around its own minimum. On the moir\'e lattice this forms an array of microscopic current loops, one on every site, all circulating the same way.

Deepening the wells compresses each pair until exchange overcomes the Zeeman energy, and the lattice crosses over to non-rotating singlets. The spin polarization and the orbital moment invert together there, the latter passing from paramagnetic rotation to diamagnetic counter-circulation.

Held polarized and taken instead to weak field, the two senses of circulation recombine into a standing wave and the atoms stop rotating. The dimers that replace the rotation are not disordered. Their axes order spontaneously into a $\sqrt3\times\sqrt3$ pinwheel crystal that breaks the three-fold symmetry of every site and triples the unit cell. The moir\'e potential cannot select that axis, since its own anisotropy is three-fold and cancels between the two lobes. The ordering comes instead from the Coulomb interaction between neighboring atoms. Pinwheel and herringbone phases are long familiar in monolayers of adsorbed molecules~\cite{Mederos1990}, but there the orienting object is a rigid molecule and nothing couples to its rotation. Here it is an electronic wavefunction, and the magnetic field couples to that rotation directly.

Both of the experiments that bear on this see part of it and not the rest. Compressibility resolves the spin of a pair but not its circulation, so the transition reported in twisted WSe$_2$/MoSe$_2$~\cite{Kometter2023} is, in the present picture, half of what happens at $V_M^*$. The orbital moment of every atom reverses, and compressibility is not sensitive to it~\cite{Foutty2025}.  Imaging resolves the shape of each atom but has so far been done at zero field~\cite{li2024wigner} where the two-lobed molecules are a product of strain~\cite{YannouleasLandman2024PRL}, not polarization, and do not spontaneously order into the pinwheel crystal that we predict at weak field.

The rotating atoms and pinwheel crystal are accessible in devices of the kind already fabricated. The field required is fixed by the moir\'e period alone, so the twist angle sets it, and a longer period brings the rotor within range of a conventional magnet. The polarization the pinwheel needs is modest at the depths where the dimers first form, reachable with a larger $g$-factor or by magnetic proximity coupling~\cite{Zhong2017,Zhao2017}. Moir\'e depth and twist angle together therefore select between a lattice of rotors, a crystal of oriented dimers, and a lattice of inert singlets.

In summary, we have characterized the collective singlet--triplet transition of rotating moir\'e atoms driven by quantized magnetic field. Resolving this competition required an ansatz with magnetic boundary conditions built in, full Landau level mixing retained and both spin sectors treated on equal footing. Neither a Landau level projection nor a mean-field treatment could accurately study the transition between these states. We 
predict a lattice of rotors, a pinwheel crystal of oriented dimers, and the singlet lattice that replaces both in deep wells. The transition into the polarized sector has already been observed~\cite{Kometter2023} but the orbital states that accompany it have not been measured and call for magnetic and imaging probes. Scanning SQUID magnetometry already resolves orbital magnetization in moir\'e devices~\cite{Tschirhart2021,Redekop2024}. The same machinery extends directly to other fillings, where the competition between the Coulomb interaction, a moir\'e potential and a magnetic field has barely been explored.

\emph{Note Added.}--- During the preparation of this manuscript, we became aware of related work by W. Zhang \textit{et al.}~\cite{Zhang2026_MagNQS} addressing a closely related problem.

\begin{acknowledgments}
\emph{Acknowledgments.}---The Flatiron Institute is a division of the Simons Foundation.
YY acknowledges support by the National Science Foundation (NSF) under grant number DMR-2532734. This research used resources of the National Energy Research Scientific Computing Center (NERSC), a Department of Energy Office of Science User Facility under Contract No. DE-AC02-05CH11231 using NERSC award DDR-ERCAP0038723 and the Hofstra Star HPC cluster, which was funded by the NSF Major Research Instrumentation (MRI) program (grant No. 2320735).
\end{acknowledgments}

\bibliography{ref}

\end{document}